\documentclass[prl,twocolumn,superscriptaddress,showpacs]{revtex4}
\usepackage[dvips]{graphicx}
\usepackage{epsfig}
\usepackage{psfig}
\usepackage{graphicx}
\usepackage{latexsym}
\usepackage[centertags]{amsmath}
\usepackage{amsthm}
\usepackage{newlfont}
\usepackage{amssymb,amsmath}

\begin{document}

\title{\textbf{SOME ASPECTS OF $2d$ INTEGRABILITY}}
\author{\textbf{M.B. Sedra} $^{2}$}
\affiliation{International Centre for Theoretical Physics, Trieste,
Italy,\\
Virtual African Center For Basic Sciences and Technology, VACBT,\\
}

\author{\textbf{A. El Boukili}}
\affiliation{\ Universit\'{e} Ibn Tofail, Facult\'{e} des Sciences,
D\'{e}partement de Physique,\\[0pt]
\ \ \ \ Laboratoire de Physique de La Mati\`ere et Rayonnement
(LPMR), K\'{e}nitra, Morocco,\\
[0pt] }

\author{\textbf{H. Erguig}}
\affiliation{\ Universit\'{e} Ibn Tofail, Facult\'{e} des Sciences,
D\'{e}partement de Physique,\\[0pt]
\ \ \ \ Laboratoire de Physique de La Mati\`ere et Rayonnement
(LPMR), K\'{e}nitra, Morocco,\\
[0pt] }
\author{\textbf{J. Zerouaoui}}
\affiliation{\ Universit\'{e} Ibn Tofail, Facult\'{e} des Sciences,
D\'{e}partement de Physique,\\[0pt]
\ \ \ \ Laboratoire de Physique de La Mati\`ere et Rayonnement
(LPMR), K\'{e}nitra, Morocco,\\
[0pt] }
\begin{abstract}
\medskip
We study the KdV and Burgers nonlinear systems and show in a
consistent way that they can be mapped to each other through a
strong requirement about their evolutions's flows to be connected.
We expect that the established mapping between these particular
systems should shed more light towards accomplishing some
unification's mechanism for KdV hierarchy's integrable systems.

\end{abstract}

\maketitle

\newpage

\section{General motivations}
An interesting subject which have been studied recently from
different point view deals with the field of non linear integrable
systems \cite{1, 2, 3, 4, 5, ss94}. These are exactly solvable
models exhibiting a very rich structure in lower
dimensions and are involved in in many areas of modern sciences and more particularly in mathematics and physics.\\

From the physics point of view, integrable systems are known to play crucial role in describing physical phenomena in many
areas such as condensed matter physics, hydrodynamics, plasma physics, high energy physics, nonlinear optics and so on.\\

Non linear integrable models, are associated to systems of non linear differential equations which can be solved exactly.
Solving such kind of differential equations in general is not an easy job, we are constrained to introduce rigorous
backgrounds such as the theory of pseudo-differential operators, Lie algebra and some physical methods such as the
scattering inverse method and the related Lax formulation \cite{1, 2}.\\

The particularity of $2d$ integrable systems is due to the
pioneering role that they deserve to the nonlinear KdV differential
equation. We focus in this work to study some properties related to
this prototype nonlinear differential equation and show how one can
reinforce its central role. We guess that such an objective is
possible since a mapping between the KdV and the Burgers's nonlinear
systems is possible by means of the Miura transformation connecting
the Lax operators of the two systems and a constrained requirements
about the associated evolution flows.
The existing mapping is expected to shed more light towards an accomplishment of the unification's mechanis\cite{Sednpb}.\\

\section{Pseudo-differential operators}
One way to introduce pseudo-differential operators \cite{ss94,
bakas, yamag} is by using the so called KP hierarchy. It's a defined
as an infinite set of differential equations. These equations are in
their turn defined through a pseud-differential operator $\cal Q$ of
the form
\begin{equation} {\cal Q}=\partial +
q_{0}\partial^{-1}+q_{1}\partial^{-2}+....
\end{equation}
where $\partial$ denotes $\frac{\partial}{\partial x}$. Another kind
of pseudo-differential operators can be obtained by using the KdV
Lax operator
\begin{equation}
{\cal L}(t)=\partial^{2} +u_{2}(x,t)
\end{equation}
with $u_{2}(x,t)\equiv u_{2}(x,t_{3},t_{5},...)$ is the KdV
potential exhibiting a conformal weight $s=2$. A typical example of
pseudo-differential operators that can emerge from ${\cal L}(t)$ is
given by its square root ${\cal L}^{\frac{1}{2}}(t)$. This is an
infinite series in inverse powers of $\partial$ namely,
\begin{equation}
{\cal
L}^{\frac{1}{2}}(t)=\partial+\frac{u}{2}\partial^{-1}-\frac{u'}{4}\partial^{-2}+(\frac{u''}{8}-\frac{u^2}{8})\partial^{-3}
+...
\end{equation}

Note that ${\cal L}^{\frac{1}{2}}(t)$ is an operator of weight
$|{\cal L}^{\frac{1}{2}}(t)|= 1$. The KdV operator ${\cal
L}(t)\equiv {\cal L}_{2}$ is also known to be the essential key
towards building the so called $2-$reduced KP hierarchy or KdV
hierarchy whose form is given by
\begin{equation}
\frac{\partial \cal L}{\partial t_{2k+1}}= [{\cal
L}_{+}^{\frac{2k+1}{2}}, {\cal L}]
\end{equation}
We have to precise that the following \emph{conventions notations}
are used:\\\\
$\bullet$ The prime derivative $u'$ is with resepect to the variable
$x$, ie $u'=\frac{\partial u}{\partial x}$. \\\\
$\bullet$ The $t_{2k+1}$ describe an infinite number of evolution
parameters associated with the KdV hierarchy. The first parameter is $t_{1}=x$.\\\\
$\bullet$ ${\cal L}_{+}^{\frac{2k+1}{2}}$ is a local operator which
means also the restriction to only positive part of the $(2k+1)$th
power of the formal pseudo-differential operator ${\cal
L}^{\frac{1}{2}}$. As an example ${\cal L}_{+}^{\frac{1}{2}}=\partial$,\\\\
$\bullet$ ${\cal L}_{+}^{\frac{2k+1}{2}}$ is an object of weight
$|{\cal L}_{+}^{\frac{2k+1}{2}}|=(2k+1) $,\\\\
$\bullet$ Given the explicit form of ${\cal L}^{\frac{1}{2}}$ one
can easily determine the form of ${\cal L}^{\frac{2k+1}{2}}$ by
proceeding as follows ${\cal L}^{\frac{2k+1}{2}}={\cal L}^{k}{\cal
L}^{\frac{1}{2}}=(\partial^{2} +u_{2})^{k}{\cal L}^{\frac{1}{2}}$

\subsection{Lax pair formalism} The principal idea, due to Lax
\cite{Lax} of this formalisms rests on our interest to solve any
given non linear system. One should emphasize that that the Lax
formalism is intimately related to the well known inverse scattering
method (ISM)\cite{Scat}. In fact given a nonlinear evolution
equation, the principal tasks is to find a linear operator whose
eigenvalues are constant under the nonlinear evolution. This is one
of the success of the ISM.
By the way, we refer the reader to \cite{2} for more details about important aspects of integrable models.\\
\subsubsection{The linear case} Let's consider a linear evolution
equation described by a time independent Hamiltonian $H$. The
question consists in finding operators whose expectation values are
preserved with time. Assume that $X$ is an operator satisfying such
a property, then from the point of view of Heisenberg picture,
$X(t)$ is required  to be unitarily equivalent to $X(0)$ such that
\begin{equation}
U^{+}(t)X(t)U(t)=X(0),
\end{equation}
where $U(t)$ is the time evolution operator given by
\begin{equation}
U(t)=exp(-iHt).
\end{equation}
Straightforward computations, based on the derivation of the last
equation from both sides, lead to
\begin{equation}
\frac{\partial X(t)}{\partial t}=i\left[X(t), H\right].
\end{equation}
Requesting for the expectation value of $X(t)$ (Its eigenvalue) to
be time independent is compatible with this equation. Furthermore,
we have
\begin{equation}
\frac{\partial U(t)}{\partial t}=iHU(t)=BU(t)
\end{equation}
where $B=-iH$ is an anti Hermitian operator $B^{+}=-B$ and
\begin{equation} U^{+}(t)U(t)=1,
\end{equation}.\\
\subsubsection{The nonlinear case} We will follow the same steps
relatives to the previous linear case and consider a nonlinear
evolution equation such that
\begin{equation}
{\cal L}(u(x,t))={\cal L}(t).
\end{equation}
denote the linear operator that we should determine with $u(x,t)$ is
the dynamical variable in $(1+1)$ dimensions. In the case of the
water waves, for example, this particular variable is interpreted as
been the height of the wave above the water surface. One can also
chow that $u(x,t)$ can exhibits a quantum number, namely the
conformal weight $s$, depending on the order of the KdV hierarchy.
For consistency requirements, one assumes that ${\cal L}(t)$ is
Hermitean and that its eigenvalues are independent of $t$. To do so,
one also suppose the existence of an unitary operator $U(t)$ such
that
\begin{equation}
U^{+}(t){\cal L}(t)U(t)={\cal L}(0).
\end{equation}
The same steps followed previously lead to the following form of the
evolution equation
\begin{equation}
\frac{\partial {\cal L}(t)}{\partial t}= \left[B(t), {\cal
L}(t)\right]
\end{equation}
One can then conclude that for ${\cal L}(t)$ to be isospectral it
must satisfy a relation similar to the linear case obtained
previously namely eq(4). The essential goal is then to find a linear
operator ${\cal L}(t)$ in $u(x,t)$ and a second one $B$, satisfying
as before $BU(t)=\frac{\partial U(t)}{\partial t}$, not necessary
linear in such a way that the commutator $[B(t), {\cal L}(t)]$
reproduces the evolution of the dynamical variable $u(x,t)$. This
means that the eigenvalues of ${\cal L}(t)$ are independent of $t$.
We have
\begin{equation}
{\cal L}(t)\psi(t)==\lambda\psi(t).
\end{equation}
where $\psi(t)=U(t)\psi(0)$ and the evolution with $t$ gives
\begin{equation}
\frac{\partial \psi(t)}{\partial t}= \frac{\partial U(t)}{\partial
t}\psi (0)=B(t)\psi(t)
\end{equation}
\textbf{ Definition:}\\ If they exist, the operators $B(t)$ and
${\cal L}(t)$ are called the Lax pair associated to a given
nonlinear evolution equation.\\
\textbf{Properties:}\\
$\bullet$ The Lax pair $B(t)$ and ${\cal L}(t)$ operators are
important in the sense that they can constitute a guarantee for
integrability of the original nonlinear evolution equation.\\
$\bullet$ The existence of $B(t)$ and ${\cal L}(t)$ means that the
nonlinear evolution equation is linearizable.\\
$\bullet$ The linear form of the evolution equation is given by
$\frac{\partial {\cal L}(t)}{\partial t}= \left[B(t), {\cal
L}(t)\right]$.\\
$\bullet$ The Lax pair $B(t)$ and ${\cal L}(t)$ play a central role
in finding the solution of the evolution equation.\\
$\bullet$ The operator $L$ is usually called the lax operator,
fixing the integrable model while $B$ is the Hamiltonian of the
system which is the local fractional power of $\cal L$.\\
\section{Towards a KdV-Hierarchy's Unification}
\subsection{The Burgers Equation}
We present in this subsection the nonlinear Burgers system.
Actually, our interest in this equation comes from its several important properties that we give as follows:\\\\
{\textbf{1}}. The Burgers equation is defined on the $(1+1)$-
dimensional space time. In the standard pseudo-differential operator
formalism, this equation is associated to the following $\cal L$-operator
\begin{equation}
{\cal L}_{Burg}=\partial+u_{1}(x,t)
\end{equation}
where the function $u_1$ is the Burgers potential of conformal
weight $|u_1|=1$. Using our convention notations\cite{ss94, Sednpb,
chinBS, ajmp, chinDES},
we can set ${\cal L}_{Burg}\in {\Sigma}^{(0,1)}_{1}$.\\
{\textbf{2}}. With respect to the previous ${\cal L}$-operator, the non linear
differential equation of the Burgers equation is given by
\begin{equation}
\dot{u}_{1}+\alpha u_{1}u'_{1}+\beta u''_{1}=0,
\end{equation}
where $\dot{u}= \partial_{t_{Burg}} u$ and $u'=\partial_x{u}$. The dimensions of the underlying objects are given
by $[t_{Burg}]=-2=-[\partial_{t_{Burg}}]$, $[x]=-1$. One can then set $t_{Burg}\equiv t_{2}$\\\\
{\textbf{3}}. On the commutative space-time, the Burgers equation can be
derived from the Navier-Stokes equation and describes real
phenomena, such as the turbulence and shock waves. In this sense,
the Burgers equation draws much attention amongst many integrable
equations.\\\\
{\textbf{4}}. It can be linearized by the Cole-Hopf transformation
\cite{hopf}. The linearized equation is the diffusion equation and
can be solved by Fourier transformation for given boundary
conditions.\\\\
{\textbf{Proposition 2}}:\\
The Burgers Lax operator ${\cal L_{Burg}}$ is a local differential operator
obtained through the following truncation of the KP pseudo-differential operator, namely
\begin{center}
${\cal L_{KP}}=\partial+u_{1}+u_{2}{\partial}^{-1}+u_{3}{\partial}^{-2}+....$
\end{center}
The local truncation is simply given by
\begin{equation}
{ \Sigma}^{(-\infty,1)}_{1}\rightarrow {\Sigma}^{(0,1)}_{1},
\end{equation}
such that to any KP pseudo operator ${\cal L_{KP}}\in { \Sigma}^{(-\infty,1)}_{1}$
\begin{equation}
{\cal L_{KP}}\mapsto \partial+u_{1}={\cal L_{Burg}}\equiv \left({\cal L}_{KP}\right)_{+}
\end{equation}\\\\
\textbf{Remark:}\\
${\cal L_{Burg}}\in {\Sigma}^{(0,1)}_{1}\equiv [{
\Sigma}^{(-\infty,1)}_{1}]_{+}\equiv {
\Sigma}^{(-\infty,1)}_{1}/{\Sigma}^{(-\infty,-1)}_{1},$
\subsection{The KdV system}
Ther KdV equation plays a central role in $2d$ integrable systems. We present here below some of
its remarkable properties\\\\
{\textbf{1}}. The KdV operator is given by
\begin{equation}
{\cal L}_{KdV}=\partial^2+u_{2}(x,t)
\end{equation}
where the function $u_2$ is the KdV potential of conformal weight $|u_2|=2$.
Using the same convention notations, we can set ${\cal L}_{KdV}\in {\Sigma}^{(0,2)}_{2}/{\Sigma_{2}^{(1, 1)}}$.\\\\
{\textbf{2}}. The Lax equation associated to the ${\cal L}_{KdV}$-operator is given by
\begin{equation}
\frac{\partial {\cal L}}{\partial t_{3}}=[{\cal L}, ({\cal L}^{\frac{3}{2}})_{+}],
\end{equation}
where straightforward calculations show that
\begin{equation}
({\cal L}^{\frac{3}{2}})_{+}={\partial}^3+\frac{3}{4}(\partial.u_{2}+u_{2}\partial)+{\cal O}(\partial ^{-1})
\end{equation}
{\textbf{3}}. Explicit form of the previous Lax equation gives
\begin{equation}
\dot{u}_{2}= 6u_{2}u'_{2}+ u'''_{2},
\end{equation}
which is nothing but the KdV equation where $\dot{u}= \partial_{t_{KdV}} u$.
The dimensions of the underlying objects are given
by $[t_{KdV}]=-3=-[\partial_t]$ and $[u_2]=2$. One can then set $t_{KdV}\equiv t_{3}$\\\\
\textbf{4}. Let's consider the KdV equation $\dot{u}_{2}= 6u_{2}u'_{2}+ u'''_{2}$. Modulo the following scalings
\begin{center}
$\partial_{t_3}\rightarrow \frac{1}{4}\partial_{t_3}$\\
$u_{2}\rightarrow \frac{u_2}{4}$
\end{center}
this equation maps to an equivalent form, namely
\begin{center}
$\dot{u}_{2}= \frac{3}{2}u_{2}u'_{2}+ u'''_{2}$
\end{center}
a form that we find in some related works (see for instance \cite{Sednpb}\\\\
\textbf{Remarks:}\\\\
\textbf{1}. For any given function $f(u)$, the action of the $\partial$-derivation on this function is given by\\
\begin{center}
$\partial.f = f'+f\partial$
\end{center}
with $f' \equiv \partial_{x} f $.\\\\
\textbf{2}. In eq(), $(\partial u_{2}+u_{2}\partial)\equiv \frac{3}{4}(u'_2+2u_2\partial)$\\\\
{\textbf{Proposition 3}}:\\
The Burgers Lax operator ${\cal L_{Burg}}$ is a local differential operator
obtained through the following truncation of the KP pseudo-differential operator, namely
\begin{center}
${\cal L_{KP}}=\partial+u_{1}+u_{2}{\partial}^{-1}+u_{3}{\partial}^{-2}+....$
\end{center}
The local truncation is simply given by
\begin{equation}
{ \Sigma}^{(-\infty,1)}_{1}\rightarrow { \Sigma}^{(0,1)}_{1},
\end{equation}
such that to any KP pseudo operator ${\cal L_{KP}}\in { \Sigma}^{(-\infty,1)}_{1}$
\begin{equation}
{\cal L_{KP}}\mapsto \partial+u_{1}={\cal L_{Burg}}\equiv \left({\cal L}_{KP}\right)_{+}
\end{equation}\\\\
\textbf{Remark:}\\
${\cal L_{Burg}}\in {\Sigma}^{(0,1)}_{1}\equiv [{
\Sigma}^{(-\infty,1)}_{1}]_{+}\equiv {
\Sigma}^{(-\infty,1)}_{1}/{\Sigma}^{(-\infty,-1)}_{1},$
\subsection{The KdV-Burgers mapping}
This section will be devoted to another significant aspect of integrable models, namely their possible unification.
In some sense, one focuses to study the possibility to establish the existence of a law allowing transitions between
known integrable systems, more notably those belonging to the generalized KdV hierarchy. The encouraging facts to follow
such a way is that these models share at least the integrability's property.\\\\
The principal focus, for the moment, is on the models discussed
previously namely the KdV and Burgers systems. The idea to connect
the two models was originated from the fact that integrability for
the KdV system is something natural due to the possibility to
connect with $2d$ conformal symmetry. We think that the strong
backgrounds of conformal symmetry can help shed more light about
integrability of the Burgers systems if one knows how to establish
such a
connection\cite{Sednpb}.\\\\
On the other hand, it is clear that these models are different due
to the fact that for KdV system the Lax operator as well as the
associated field $u_{2}(x,t)$ are of conformal weights $2$, whereas
for the Burgers system,
${\cal L}_{Burg}$ and $u_1$ are of weight $1$.\\\\
Our goal is to study the possibility of transition between the
two spaces ${\Sigma}_{2}^{(0,2)}/{\Sigma}_{2}^{(1,1)}$ and ${\Sigma}_{1}^{(0,1)}$
corresponding respectively to the two models.  This transition,
once it exists, should lead to extract more information on these models and also on their integrability.\\\\ To
start, let us consider the following property
\\\\
{\textbf{Proposition 4}:}\\ Let us consider the Burgers Lax operator
$L_{Burg}(u_1)=\partial + u_{1}\in {\Sigma}_{1}^{(0,1)}$. For any
given $sl_2$ KdV operator ${\cal L}_{KdV}(u_2)={\partial}^2+u_{2}$
belongings to the space ${\Sigma}_{2}^{(0,2)}/{\Sigma}_{2}^{(1,1)}$,
one can define the following mapping
\begin{equation}
{\Sigma}_{1}^{(0,1)} \hookrightarrow {\Sigma}_{2}^{(0,2)}/{\Sigma}_{2}^{(1,1)},
\end{equation}
in such a way that
\begin{equation}
{\cal L}_{Burg}(u_1)\rightarrow {\cal L}_{KdV}(u_2)\equiv
{\cal L}_{Burg}(u_1)\times {\cal L}_{Burg}(-u_1).
\end{equation}\\
We know that the space ${\Sigma}_{2}^{(0,2)}$ of KdV Lax
operators of weight $s=2$ is different from the one of operators of weight $s=1$ namely ${\Sigma}_{1}^{(0,1)}$.
What we are assuming in this proposition is
a strong constraint leading to connect the two spaces. This
constraint is also equivalent to set
\begin{equation}
{\Sigma}_{2}^{(0,2)}/{\Sigma}_{2}^{(1,1)}\equiv {\Sigma}_{1}^{(0,1)}\otimes { \Sigma}_{1}^{(0,1)}
\end{equation}
Next we are interested in discovering the crucial key behind the
previous proposition. For this reason, we underline that this
mapping is easy to highlight through the well known Miura transformation
\begin{equation}
L_{KdV}={\partial}^2+u_{2}=(\partial^1+u_{1})\times(\partial^{1}-u_{1})
\end{equation}
giving rise to
\begin{equation}
u_{2}=-u_{1}^{2}-u'_{1}.
\end{equation}
This is an important property since one has the possibility to
realize the KdV $sl_2$ field $u_2$ in terms of the
Burgers field $u_1$ and its derivative $u'_{1}$. This realization shows among other an
underlying nonlinear behavior in the KdV field
$u_{2}$ given by the quadratic term $u^{2}_{1}$.
\\\\However, \emph{proposition 1} can have a complete and
consistent significance only if one manages to establish a
connection between the differential equations
associated to the two systems.  Arriving at this stage, note that
besides the principal difference due to conformal weight, we stress
that the two nonlinear evolutions equations: ($\dot u_2=\frac{3}{2}uu'+\theta^{2}u''')$ and $\frac{1}{2\theta}\frac{\partial u_1}{\partial
t_{2}}+2(1-\eta)uu'-\xi u''=0$ of KdV and Burgers systems respectively are distinct by a
remarkable fact that is the KdV flow $t_{KdV} \equiv t_3$ and the Burgers one $t_{Burgers}\equiv t_2$ have
different conformal weights: $[t_{KdV}]=-3$ whereas $[t_{Burgers}]=-2$.
\\\\In order to be consistent with the objective of \emph{proposition 1}, based on the idea of the possible
link between the two integrable systems, presently we
are constrained to circumvent the effect of proper aspects
specific to both equations and consider the following second
property:
\\\\
{\textbf{Proposition 5}}\\\\
By virtue of the Burgers-KdV mapping and dimensional arguments,
the associated flows are related through the following ansatz
\begin{equation}
\partial_{t_2}\hookrightarrow
\partial_{t_3}\equiv\partial_{t_2}.\partial_{x}+\alpha
\partial_{x}^{3}
\end{equation}
for an arbitrary parameter $\alpha$.
\\\\With respect to assumption (), relating the two
evolution derivatives $\partial_{t_2}$ and $\partial_{t_3}$
belonging to Burgers and KdV's hierarchies respectively, one should
expect some strong constraint on the Burgers and KdV currents.\\
We have to identify the following three differential equations
\begin{equation}
\begin{array}{lcl}
\partial_{t_3}u_2&=& 6u_{2}u'_{2}+ u'''_{2},\\\\
&=&-2u_{1}\partial_{t_3}u_{1}-\partial_{t_3}u'_{1},\\\\
&=&\partial_{t_2}u'_{2}+\alpha u'''_{2}.
\end{array}
\end{equation}
{\bf Acknowledgements} \\
MBS would like to thank the Abdus Salam International Center for
Theoretical Physics (ICTP) for hospitality and acknowledge the
considerable help of the High Energy Section and its head Seif
Randjbar-Daemi for the invitation. Acknowledgements to OEA-ICTP and
to its head George Thompson for valuable scientific helps in the
context of Net-$62$.

\end{document}